\documentstyle[12pt]{article}
\input{psfig}
\begin{document}
\baselineskip 23pt
\bibliographystyle{unsrt}
\pagestyle{plain}
\title{Generic Modeling of Chemotactic Based Self-Wiring of Neural
  Networks} 
\author{Ronen Segev and Eshel Ben-Jacob \( ^* \)  \\
School of physics and astronomy,\\
 Raymond \& Beverly Sackler Faculty of exact Sciences,\\
Tel-Aviv University, Tel-Aviv 69978, Israel \\
\(*\) communicating author.\\}

\maketitle







\begin{abstract}
The proper functioning of the nervous system depends critically on the
intricate network of synaptic connections that are 
generated during the system development.
During the network formation, the growth cones
migrate through the 
embryonic environment to their targets using chemical
communication. A major obstacle in the elucidation of fundamental
principles underlying this 
self-wiring is the complexity of the system being analyzed.
Hence much effort is devoted to in-vitro experiments of simpler 2D
model systems.  In these experiments neurons are 
placed on Poly-L-Lysine (PLL) surfaces so it is easier to monitor
their self-wiring. 
We developed a model to reproduce the salient features of the
2D systems, inspired by the study of bacterial colony's growth and the
aggregation of amoebae.    
We represent the neurons (each composed of cell's soma,
neurites and growth cones) by active elements that capture the
generic features of the real neurons.
The model also incorporates stationary units representing the cells'
soma and communicating walkers representing the growth cones. 
The stationary units send neurites one at a time, and respond to
chemical signaling. 
The walkers migrate in response to chemotaxis substances emitted by
the soma and 
communicate with each other and with the soma by means of chemotactic
``feedback''.  
The interplay between the chemo-repulsive and chemo-attractive
responses is determined by the dynamics 
of the walker's internal energy which is controlled by the soma.
These features enable the neurons to perform the complex task
of self-wiring.
We present numerical experiments of the model to demonstrate its
ability to form fine structures in simple networks of few neurons.
Our results
raise two fundamental issues: 1. One needs to develop characterization
methods (beyond number of connections per neuron) to distinguish the
various possible networks. 2. What are the relations between the
network organization and its computational properties and efficiency. 

\end{abstract}

\section{Introduction}

The brain is probably the most challenging complex system
that scientists can study \cite{Abeles91}. And indeed, much effort has been
devoted to brain  
studies from the physiological level of ionic channels to the
philosophical level where 
questions about intelligence and self-awareness 
are discussed \cite{Penrose}.
Yet, one of the 
fascinating aspects about the brain has almost been completely ignored until
recently. We refer to the process during which a collection
of individual neurons are transformed into a functioning network with unique 
capabilities-the brain. This emergence process cannot be totally
determined by the 
stored genetic information. In a human brain, for instance, there are 
approximately $10^{11}$ neurons that form a network with more than $10^{15}$
synaptic connections. The human $DNA$ is composed of about $10^9$ bases, so it 
lacks sufficient memory storage of the detailed structure of a brain.
The alternative extreme explanation, of total randomness, could not
be correct as well. After all, we know that while on the micro level (up to
about $1mm$) the structure appears to be random, on the macro level
(above $1cm$) the brain's structure is quite deterministic.
In addition, the brain structure varies from species to 
species, and, within a given species, some brain skills are inheritable
(vs. learned). So clearly some elements of the brain structure must
be dictated by stored genetic information. 

The contemporary view is that the brain structure is essentially
deterministic on a large scale but probabilistic on a small scale
\cite{Abeles91}. As 
a consequence the network has no optimal structure. But we believe that
neural networks can be constructed in an optimal way, and this optimal
way is derived from the biological mechanisms that construct the
network. In any approach to the construction of the network there
must be a clear strategy by which the neurons find each other
to establish synaptic connections. 

At present we do not know which parts of the 
information about the brain structure are stored, and the general
question about the role of randomness vs. determinism during the
emergent process is still open \cite{Abeles91}. 

Neuronal connections are formed when each neuron sends neurites
that migrate through the embryonic
environment \cite{TL96}. 
At the leading tip of each neurite there is a region called
growth cone, which is capable of ``measuring'' concentrations and
gradients of chemical fields. Indeed, the growth cones
navigate using sophisticated means of chemical signaling for
communication and regulation, including repulsive and attractive
chemotaxis.

At the beginning of the      
growth process the neurite has to migrate from its own cell's
soma. The neurite migrates to the area in which it is supposed to form
a synaptic    
connection. In this area there are many neurons, each of which is a
possible target cell for the neurite.
When the neurite approaches one of the possible
target cells, with which it will finally form a synaptic connection, 
it has to be attracted to that cell's soma. 

We describe here elements of a
novel strategy for the emergent process. Since
the proposed model is  
for simplified $2D$ systems, as such, it is far from being a full
description 
of the brain's adaptive self-wiring. Yet, if tested and shown to be
correct, it will  
provide an important step towards understanding the emergence process
in a real brain.

To clarify the problem at hand we show 
in figure \ref{culture} a self wiring process of neurons on a 2D
silicone 
surface. Since we cannot distinguish between dendrites and axons,
we refer to both as neurites from here on.
The thin filaments are the neurites and during development
they migrate over the surface to form synaptic connections. 

Our model of self-wiring is inspired by communicating walkers model
used in the study of complex patterning of bacterial colonies.
We have developed \cite{SB98,SB98a} 
a simple model to reproduce the salient features of the 2D systems. 
We represent the neurons (each composed of cell's soma,
neurites and growth cones) by simple active elements that capture the
generic features of the real neurons. 
The model also incorporates stationary units representing the cells'
soma and communicating  
walkers representing the growth cones. 
The stationary units send neurites one at a time, and respond to
chemical signaling. 
The walkers migrate in response to chemotaxis substances emitted by
the soma and 
communicate with each other and with the soma by means of chemotactic
``feedback''.  
The interplay between the chemo-repulsive and chemo-attractive
responses 
is determined by the dynamics
of the walker's internal energy which is controlled by the soma.
These simple features enable the neurons to perform the complex task of
self-wiring. In section 3 we elaborate on this model and its
experimental predictions.

\section{Biological and Experimental Background}

In this section we present a brief summary of the existing biological
knowledge relevant to the self-wiring of neural nets and their
modeling. 

\subsection{Cell cultures}

There are two major types of cell cultures used as a model (note that
we use in this section a different meaning to the word ``model'') for
neuronal 
development: Primary cultures, which are prepared from cells taken
directly from animal (usually rat embryo), and neuron-like cultures
of cell lines. A cell line is usually derived from tumor cells
that were cloned so as to obtain a genetically homogeneous
population. Initially, the cell line has no particular biological
function, instead, upon exposure to an external trigger (for
example exposure to nerve growth factor (NGF) for PC12 cell line),
it is pushed to develop neuronal-like processes. It develops morphological
and functional neuronal phenotypes such as extending neurites,
manufacturing neuro-transmitters and becoming electrically
excitable. However, it is not clear how far these cells go in the route
of neuronal differentiation (For example the PC12 cell line does not
form real synapses). In figure 1 we show cell culture of
neurobalstoma cells, which develop neuron-like processes when they are
exposed to a serum free media.

\subsection{Growth cones: structure and dynamics}

The growth cone is capable of measuring concentration and
concentration gradients of substances in the environment. 
It is composed of a central core which is an extension of
the neurite itself, and is rich with microtubulets that provide the structural
support. The core is rich with mitochondria, endoplasmic reticulum and
vesicular structures. Surrounding the central core are regions known as
lamellipodia, in which the contractile protein Actin is abundant. At
the extremities of the lamellipodia there are very thin straight
filaments known as filopodia. The filopodia are in constant motion, as
they extend from the lamellipodia and retract back to it.
The growth of the neurite occurs when filopodia extend from the
lamellipodia and 
remains extended rather than  retracts as the end of the lamellipodia advances
towards the filopodia.
The complexity and the dynamics of the growth cones hint that they
might act as autonomous entities. Indeed, there are direct
experimental observations of 
the activity of growth cones that support this view \cite{HM84}.
In the experiment growth cones are cut off from their neurite. These
isolated growth 
cones continue to extend and/or react to chemicals for a considerable
long time after they were severed from their cells' soma.
This observation indicates that the measurements of chemotaxis
gradients 
occur locally through receptors on the growth cones themselves, and do
not require any signal from the cell's soma.  

The above observations are essential to the construction of our model.
In particular, they lead us to represent each of the growth cones as an
independent entity (walker) with its own internal energy as described
in the next section. 

Microscope observation \cite{BR97} reveals that the movement of the
growth 
cones appears to be a non-uniform random walk which has the highest
probability to move forward (``inertia'') and high probability to move
backward (``retraction''). 
The movement is not continuous, as there are time intervals during
which the growth cones do not move. 
The growth rate is of the order of micrometer per minute \cite{GB79}. 

\subsection{Chemotaxis}

Extensive studies in-vivo and in-vitro 
revealed that the movement of the growth cones can be affected by four types
of guidance cues: attractive or repulsive cues  
that can be either local or non-local. The local cues are
contact mediated by non-diffusive cell  
adhesion molecules (CAM) and extra cellular matrix (ECM) molecules. 
The non-local forces are mediated by  
emission of chemo-attractant and chemo-repellent substances which
``pull'' 
and ``push'' the growth cone from the soma or its neurites \cite{TL96}.

While the existence of attractive and repulsive agents has been
established long ago, there recently has been an experiment which
demonstrated the existence of a triggering agent. 
A gradient of brain-derived neuro-tropic
factor (BDNF) was created with a micro-pipette near a growth cone 
\cite{SMP97}. The specific kind of growth cone used in the
experiment is usually attracted by the gradient of the BDNF. 
But the same gradient of BDNF induced a repulsive response of the
growth 
cones when the growth cones were cultured in the 
presence of a competitive analogue of $cAMP$ in the media. 
Apparently, the analogue of $cAMP$ can trigger the growth cones to
react differently to the same chemotaxis substance. 

\subsection{Chemical perception}

Biological elements (growth cones, cells) sense the local concentration
of a chemical via membrane receptors binding the chemicals molecules
\cite{EBCP97}. They sense the concentration by measuring the fraction of
occupied receptors, $N_0/(N_f+N_0)$, where $N_0$ and $N_f$ are the number
of occupied and free receptors respectively. For a given concentration $C$,
$N_0$ is determined by two characteristic times: the average time the
receptor is occupied, $\tau_0$, and the average time the receptor is free,
$\tau_f$. Since $\tau_f \propto 1/C$ with the proportion coefficient, $k$,
determined by the receptor affinity to the chemical, we get:
\begin{equation}
\frac{N_0}{N_f+N_0}=\frac{\tau_o}{\tau_f+\tau_0}=\frac{C}{K+C}
\end{equation}
where $K=k/\tau_0$. Since growth cones measure changes in $N_0/(N_0+N_f)$
and not in the concentration itself using eq (1) and assuming $\tau_0$
does not change in space we obtain:
\begin{equation}
\frac{\partial}{\partial x}\left(\frac{N_0}{N_0+N_f}\right)=\frac{K}{(K+C)^2}
\frac{\partial C}{\partial x}
\end{equation}
As we can see the growth cone measures the chemical gradient
multiplied by a
prefactor $K/(K+C)^2$. This law is known as the ``Receptor's Law''.
Since at very high concentrations all the receptors
are occupied the response is zero. At very low concentrations, due to
internal and external noise, the response vanishes as well. 

\subsection{Constrains on the navigation distance}

There are two related constrains which the distribution of a diffusible
factor must satisfy to provide an effective guidance cue to a specific
location \cite {GH98}. First, as was explained in the previous
section,  
the absolute concentration of the chemotaxis agent must not be 
too small or too large. Second, the gradient across the
growth cone must be
sufficiently large, since the growth cone measures the
concentration differences over its width. These constrains are related
because in both cases the problem is to overcome the statistical
noise.  
Goodhill \cite {GH98} investigated the limitation these constrains
impose on the maximum
guidance range of a diffusible factor, emitted by a cell away from the
growth cone, by estimating the diffusion
constant, the rate of production of a chemotaxis agent, the minimum
and maximum of concentration and gradient detection. He came to the
conclusion that maximum guidance distance may range up to $1mm$.

\subsection{Synaptic connections}

An extending neurite
does not make synaptic contacts with every cell that it encounters on
its path, there has to be a signal which instructs the growth cone to
slow its growth, contact one of the possible target cell, and
form the synaptic connection. The nature of this signal is as yet
unknown, but it is a  
reasonable assumption that it is related to some chemical
interaction between the growth cone and the target cell. We did not
include such interaction in our model, we simply assumed that the
growth cone makes a synaptic connection when it first reaches a cell's
soma. 

\subsection{Spontaneous release of transmitter from growth cones}

There is an experimental evidence for spontaneous release of
neuro-transmitter acetylcholine from growth cones. The release of
material from a growth cone may have a role in 
the interaction between the growth cone and its immediate
environment. This observation led as to assume the existence of a
chemical agent that the growth cones use to communicate with the
target cells as we describe below.

\section{The self-wiring model}

Our model of self wiring is inspired by the communicating
walkers model used in the study of complex patterning of bacterial
colonies \cite{EB94,EB95}, and by the bions model used in the study
of amoebae aggregation \cite{KL93}. 
How should one approach modeling of the growth of neural network?
With present computational power it is natural to use computer models
as 
a main tool in the study of complex systems. However, one must be
careful not to be trapped in the ``reminiscence syndrome'', described
by 
J. Horgan \cite{Horgan95}, as the tendency to devise a
set of rules which will mimic some aspects of the observed phenomena
and then, to quote J. Horgan {\it``They say: `Look, isn't this
reminiscent of a biological or physical phenomenon!' They jump in
right away as if it's a decent model for the phenomenon, and usually 
of course it's just got some accidental features that make it look 
like something.''} Yet, the reminiscence modeling approach has some
indirect value. True, doing so does not reveal directly the
biological functions and behavior. However, it does reflect
understanding of geometrical and temporal features of the patterns,
which indirectly might help in revealing the underlying biological
principles. 

Another extreme is the ``realistic modeling'' approach,
where one constructs an algorithm that includes in detail all the
known biological facts about the system. Such an approach sets a
trajectory of including more and more details vs. generalized
features. The model keeps evolving to include so many details that it
loses any predictive power.

Here we adopt another approach called ``generic modeling''
\cite{KL93,EB94}, where we seek to elicit, from the
experimental observations and the biological knowledge, the generic
features and basic principles needed to explain the biological
behavior and to include these features in the model. We will
demonstrate that such modeling, with close comparison to experimental
observations, can be used as a research tool to reveal new
understanding of the biological systems. 

\subsection{The model overview}

In modeling the neurite navigation we were inspired by the bions
model used in the study of amoebae aggregation \cite{KL93} and by the
communicating 
walker model used in the study of bacterial colonies
\cite{EB94} \cite{EB95} \cite{EBCP97}. 
In the model the growth cones are represented
by walkers which perform off-lattice non-uniform random
(biased) walk \cite{SB98}. The chemical dynamics (e.g chemotactic
agents, 
triggering field) are described by continuous
reaction-diffusion equations solved on a tridiagonal lattice with a
lattice constant $a_0=10\mu m$.
Each of the soma is represented by a stationary (not moving) unit
occupying one lattice cell. The neurite are simply defined as the
trajectory performed by the growth cone.

We assume that each of the soma cells continuously emits a repulsive
agent whose concentration is denoted by $R$. In the model, $R$
satisfies the following reaction diffusion equation:
\begin{equation}
   \frac{ \partial R}{ \partial t}=D_R \nabla^2 R +
   \Gamma_{R} \sum_{{ }^{soma}} \delta ( \vec{r} - \vec{r_j} )
   -\lambda_R R
\label{repulsive-field}
\end{equation}
Where $D_R$ is the diffusion coefficient, $\lambda_R$ is the
spontaneous decomposition rate and $\Gamma_R$ is the emission rate by
the soma cells.  

When a neurite first sprouts it is mainly affected by the repulsive
agent 
and moves away from its ``mother'' soma cell. It then continues to
move on a trajectory which maximizes the distances from the
surroundings soma cells (fig \ref{drawing}a).

When a neurite reaches a specific length 
determined by its soma cell via the internal energy
(in a manners described below), it does two things: 1. It switches
its chemotactic sensitivity from sensitivity to the repulsive
agent to sensitivity to the attractive. 2. It emits a quantum
of a triggering material (which satisfies an equation similar to
eq. \ref{repulsive-field}). Soma cells in the neighborhood respond by
emitting a quantum of
attractive agent if they sense an above threshold concentration of the
triggering material. As a result, the growth cone moves towards the
soma cell with the strongest attractive response (typically, the one
closest to the growth cone, see figure \ref{drawing}b)

\subsection{The growth cone's movement in the absence of chemotaxis}

At the absence of chemotaxis each walker performs off-lattice random
walk of step size $d$ which is of the order $.5a_0$, where
$a_0\sim 5\mu m$ is the lattice constant.
The random walk is at an angle \( \theta \) which is chosen out of
12 available directions \( \Phi{(n)} = (n-1)\pi /6 \). The angle
$\phi(n)$ is measured relative to the last direction of movement
($n=0$ corresponds to a step in the last direction of movement).
The angle is chosen from a non-uniform
probability distribution \( P_o (n) \) shown in Fig 4. 
The highest probability is to continue to move in the same
direction, the second pick in the probability distribution is to
move backward.  
Thus the walker moves from its location \( \vec{r_i} \) to a new location 
$\vec{r'_i}$ 
given by :
\begin{equation}
\vec{r'_i}=\vec{r_i}+d(cos \theta,sin\theta)
\end{equation}

A typical trajectory of such a random walker is demonstrated in figure
\ref{noise2}. The hexagon represents the cell's soma and the dot
represents the 
walker. This random walker movement generates a neurite shape that
is not in agreement with experiments \cite{Dwir96}. 
Thus we modify the motion rule as follows:
the walker does not move every time step. After \( \theta \)
is selected, 
a counter for that chosen direction (given n) is increased by one. 
The walker performs a movement only after one of the counters reaches
a specified 
threshold \( N_C \). The movement is in the direction which
corresponds  
to that counter. This process acts as a noise reduction mechanism. We
show a typical trajectory of the walker in figure \ref{noise1} The observed
neurite shape is in better agreement with experiments \cite{Dwir96}.

\subsection{Growth cone movement in the presence of chemotaxis}

In the presence of chemotactic materials, the probability distribution
\( P_0 (n) \) (the 
relative probability of choosing from the available 12 directions) is
modified. Since a growth cone is repelled (attracted) by the repulsive
(attractive) agent, the
probability is higher (lower) in the direction of low (high)
directional derivative of the chemo-repellent agent's concentration. 
In our model we assumed that the probability changes linearly 
in the chemo-repellent concentration gradient. 
A similar rule is used for the chemo-attractant concentration.   
The new probability of moving in the n-th direction is given by:
\begin{equation}
P(n) = P_0 (n) + G_A \cdot S(A) \nabla_n A - G_R \cdot S(R) \nabla_n R
\end{equation}

Where \( A \) and \( R \) are the concentrations of chemo-attractant
and chemo-repellent,  respectively. 
\( \nabla_n \) is the directional derivative in the appropriate
direction. 
The functions \( S(A) \) and \( S(R) \) are pre-factors which describe
the receptor's sensitivity as a function of the concentration (the
``Receptor's law'' as described in section 2.4).
\( G_R \) and \( G_A \) are functions of $\frac{\mbox{d}E}{\mbox{d}t}$.
They determine the relative magnitude of response to 
chemo-attractant and the chemo-repellent. 
Since \( \frac {\mbox{d}E_i}{\mbox{d}t} \) decreases with the neurites
length (as we describe in section \ref{internal-energy}), 
and we assume that the growth cone is more sensitive to the
chemo-repellent while it is close to its cell's soma, 
\( G_R \) is taken to be an increasing function of 
\( \frac{\mbox{d}E_i}{\mbox{d}t} \) as described in figure
\ref{functions}a, to obtain 
the desired result. A similar rule is used for the chemo-attractive
concentration. Since the walker is more sensitive to the chemo-attractant
while it is far away from its soma we take $G_A$ to be a decreasing
function of $\frac{\mbox{d}E_i}{\mbox{d}t}$ as described in figure \ref{functions}b.

\subsection{Chemicals concentration}

We handle the corresponding continuous reaction-diffusion equations
for the chemical concentrations by solving them on a triangular lattice.
The equations for the chemo-repellent concentration R, chemo-attractant
A and triggering T are given by : 
\begin{equation}
   \frac{ \partial R}{ \partial t}=D_R \nabla^2 R - \lambda_R R + \Gamma_{R} \sum_{{ }^{soma}}
\delta ( \vec{r} - \vec{r_j} )
\end{equation}
\begin{equation} 
   \frac{ \partial A}{ \partial t}=D_A \nabla^2 A - \lambda_A A + \Gamma_{As} \sum_{{ }^{emitting}_{soma}} \delta ( \vec{r} - \vec{r_j} )
+\Gamma_{Agc} \sum_{{ }^{emitting}_{walkers}} \delta ( \vec{r} - \vec{r_i} )
\end{equation}
\begin{equation} 
   \frac{ \partial T}{ \partial t}=D_T \nabla^2 T - \lambda_T T + \Gamma_{T} \sum_{{ }^{emitting}_{walkers}}
\delta ( \vec{r} - \vec{r_i} )
\end{equation}
where \( D_R,D_A,D_T \) are the diffusion coefficient,
$\lambda_R$,$\lambda_A$ and $\lambda_T$ are the rate of spontaneous
decomposition of  
\( R,A,T \) respectively, \( \Gamma_R \) is the rate of emission of
R by the soma located at \( \vec{R_j} \). $\Gamma_{As}$ and
$\Gamma_{Agc}$ are the rate of emission of A by the soma and the
growth cone 
respectively and $\Gamma_{T}$ is the rate of emission of triggering
agent by a growth cone located at $\vec{R_i}$. The summation in
equation 3 is over all the soma. The summation in equations 4 is over
all the soma and walkers that are currently emitting chemo-attractant.
We assume that \( D_R \) and \( D_A \) are of the same
order. We further assume that \( \lambda_{R} < \lambda_A \), so the
chemo-repellent is long ranged with respect to the chemo-attractant.  

\subsection{The growth cone's internal energy}
\label{internal-energy}

The assignment of internal energy to describe the metabolic state of
the growth cone is a crucial assumption in 
the model. The assumption was first motivated by the modeling of
bacterial colonies \cite{EB94,EB95} \cite{MA97} in which such   
internal energy turned out to be a crucial feature in modeling systems
composed of biological elements. 

Since we assume that the walker changes its sensitivity to the
chemo-repellent and the chemo-attractant as a function of its neurite
length, the walker can migrate away from
its cell's soma at the beginning of the growth and attract to one of the
target cell's soma 
at the late stage of the growth. We propose that the dependence of
the walker sensitivity on the neurite length enters via an energy
function which depends on the neurite length. The walker changes
its relative sensitivity to the chemotaxis as a function of the energy
as we describe below.
This assumption is supported by the experimental observations that the
growth cones are rich with mitochondria \cite{TN91}. 
We assume that the soma feed the 
growth cone with internal energy, which the growth cone utilizes for
its metabolic processes. We further assume that 
the neurite consumes internal energy proportional to its length. 
Finally, it is natural that the growth cone spends internal
energy at a constant rate for its metabolic process.
Thus, the time evolution of the internal energy is given by : 
\begin{equation} 
\frac{\mbox{d}E_i}{\mbox{d}t}=\Gamma(N_j)-\Omega-\lambda L_i +K(A)A
\end{equation}
Where \( \Gamma (N_j) \) is the rate of internal energy supplied by the
soma, it is a decreasing function of \( N_j \), the number of neurites
sent out by the 
soma. The growth cone consumes internal energy at a rate \( \Omega \), 
and its neurite consumes the internal energy at a rate \( \lambda \)
per 
unit length. The last term on the right hand side of eq. (4) describes
the 
absorption of chemo-attractant by the growth cone. $A$ is the
chemo-attractant concentration at the walker position and $K(A)$ is a
absorption coefficient.
We assume (as is usually the case \cite{TL96}) that the
chemo-attractant agent can be used by the growth cone as an energy source. \( K(A) \) is already measured in units of energy.

\subsection{Walker activity}

The soma supplies energy at a higher rate than the walker's
consumption rate \( \Omega < \Gamma (N_\j) \) (equation 4).  
Hence initially (i.e. short neurite's length \( L_i \)) the internal
energy 
increases. At this stage of the growth the walker has to migrate away
from its own soma, therefore we assume that the walker is
very sensitive to the chemo-repellent and insensitive to the
chemo-attractant when $\mbox{d}E/\mbox{d}t$ is positive. 

In order to overcome the repulsive force and 
attract to one of the possible target cells at the late stage of the
growth, we assume two assumptions: the walker reverts its response to
the chemotaxis, and signals to the target cell about its presence
when $\mbox{d}E/\mbox{d}t<0$.  
Specifically we can write in the absence of chemo-attractant, for 
\begin{equation}
\begin{array}{rl}
L_{i} > L_{c}  & \mbox{where } L_c\equiv (\Gamma - \Omega )/ \lambda  
\end{array}
\label{critical-length}
\end{equation}
the internal energy decreases. When $\frac{\mbox{d} E_i}{\mbox{d} t}$
first becomes negative, the walker
emits a quantum of triggering material. The triggering substance
diffuses through the media and is sensed by a possible target cell's
soma. The 
target cell's soma, as we described above, react by emitting a quantum of
attractive substance. The walker can find its way to the target cell's
soma using this attractive substance concentration gradient.
We call this signaling mechanism of the walker to the target cell an
``Attractive Chemotactic Feedback''.
The walker response to the chemotaxis changes during the growth
process in a continuous manner as described in the next section.

After the walker emits the first quantum of triggering agent it waits
$\tau_T$ time units before the next quantum is emitted.
If during \( \tau_T \) the growth cone
senses sufficient concentration of chemo-attractant, or 
\( \frac{\mbox{d}E_i}{\mbox{d}t}\) becomes positive, it will not emit
another quantum of the triggering 
agent. If \( \frac{\mbox{d}E_i}{\mbox{d}t} \) is negative for a
sufficiently long time, so that 
\( E_i \) drops to zero, the neurite and its growth cone degenerate and
 are removed, as usually is the case for growth cones that fail to
create an appropriate synaptic connection. 
When the walker reaches another cell or
another cell's neurite, it creates a synaptic connection and its
metabolic processes are stopped.

\section{Simulation of The Chemotactic Navigation}

As was mentioned above the reaction-diffusion equations are solved on a
tridiagonal lattice with a  
lattice constant \( a_0 \). Thus, the fact that
the soma occupies one lattice cell is in   
agreement with the typical size of the neurons' soma. The typical  
size of the simulated system is about \( 200-400 \) \( a_0 \) and the
distance between cells' soma is about $25a_0$. A typical
diffusion coefficient \( D \) of the chemicals is of the 
order of \( 10^{-6}-10^{-7} cm^2/sec \). Time is measured in units of
$1sec$ and the dimensionless diffusion 
coefficients are of the order of 10. 
In the simulations, the walkers growth rate is about $0.01$ in
dimensionless units, which corresponds to $\sim 1 \mu/min$ in
agreement with the measured advance rate of the growth cones. 

In figure \ref{normal} we show the result of numerical simulation of
self wiring of 50 cells system on a $200\times200$ grid. 
We look for new experiments that will provide a crucial test 
to verify our model's validity and its ability to form
the delicate structure of connection.

\subsection{Two-cell systems}

To demonstrate the efficiency of the proposed chemotactic navigation
strategy, we consider two-cell systems in various
configurations. In figure \ref{twocells}, the 
synaptic connections are formed at about half-way between the cell's
soma. The wiring process is very efficient: five out of the
six emitted walkers formed connections.

In figure \ref{twocellvariance} the cell on
the right is a ``normal'' cell, while the one on the left is a
``variance'' which is incapable of emitting neurites. This choice is
in order to make the wiring pattern more 
transparent. We see that even neurites which originally migrated
away from the target cell change their path and migrate towards this
cell, once the target cell is triggered to emit a chemo-attractant. 

In figure \ref{barrier} we consider a system as in figure
\ref{twocellvariance} with a
barrier in between, in the absence of chemotactic communication (figure
\ref{barrier}a), the 
barrier prevents the formation of synaptic connections between 
the two cells. When included, the chemotactic communication enables
the walkers to
over come the barrier effect and the two cells are wired as is shown
in figure \ref{barrier}b.

In figure \ref{saturated} we show the effect of a saturated media
with chemo-attractant on the growth process.  
The media in which the growth process takes place is enriched
artificially with chemo-attractant agent. In our model we assume that
the chemo-attractant feeds the walkers (the $K(A)A$ term in equation
3.6), therefore, in chemo-attractant rich media the internal energy time
derivative remains at a high level for a long time.
Since $\mbox{d}E/\mbox{d}t$ remains positive
the growth cones do not emit the triggering agent, and the
target cell does not emit chemo-attractant. As a result, the growth
cones cannot find their path to the target cell. This computer
simulation suggests an experiment to prove the validity of our   
assumption concerning the existence of internal energy. We predict
that in an artificially enriched chemo-attractant media the growth cones
will lose their ability to find their path.

\subsection{The effect of cutting off the growth cone}

According to our model when
a growth cone is cut off from its soma, the growth cone will change 
its response to the chemotaxis. 
This observation give us an indirect way to
test the existence of internal degree of freedom in the growth.
\cite{SMP97,HM84} \cite{change_response98}). 

\subsection{Growth regulation via internal energy}

In fig 12 we show a system of five
cells: one normal, at the center, and four 
"variance" cells at the corners. All the parameters in figures
\ref{5cell}a and \ref{5cell}b 
are the same, but in figure \ref{5cell}b the cell at the center
``feeds'' the 
neurite at a higher rate. As a result, the growth cones trigger target
cells when they are further away from their soma, and the soma is
wired to all four neighbors and not only to two, as is the case in
figure \ref{5cell}a. 

The fact that the feeding rate has a dramatic effect on the efficiency
of the wiring process brings to mind the idea that the soma can
regulate the distance of connections by adjusting the feeding rate. In
figure \ref{30cell} we show the same system as in figure \ref{5cell}
but with additional 24 ``variance'' cells. The first two
neurites are fed at a low rate. Hence they are connected to the
nearest neighbors. The third neurite is fed at a higher rate. Thus it
forms a connection further away. 

\subsection{Simulation of micro-pipette as a source of chemo-attractant}

A direct way to prove that a Nerve Growth Factor (NGF) is a chemotaxis
is to place a micro-pipette containing NGF near a growing axon in tissue
culture, the axon turns and grows toward the NGF source 
\cite{GB79}. This turning response to elevated concentrations
of NGF represent the chemotactic guidance of the NGF.
In our model, we assume that at the beginning of the growth process
the growth cones are insensitive to the chemo-attractant, here we
construct a similar experiment to show the validity of our assumption.

The model predicts that since the walker has low sensitivity to
chemo-attractant at the 
beginning of the growth process, an artificial source of
chemo-attractant will not affect the growth at early stages. We show a
computer 
simulation of such a system in figure \ref{micro-pipette}. We
simulated one cell in the 
presence of an artificial source of chemo-attractant and the contours
represent the concentration levels of the chemo-attractant. We can see
that the walker's growth is not affected from the presence of
chemotaxis gradient.

\subsection{The growth in the presence of 2-fold anisotropy}

Recently, there have been experimental studies of the effect of
imposed 2-fold anisotropy on the wiring process \cite{Dwir96}.
In the experiments, Dwir el al cultured hundreds of neurons on a
silicon wafer covered with Poly-L-Lysine stripes, along which the
growth cones have higher probability to move.
To mimic the imposed anisotropy
we include in the model a comb of stripes ,$5a_0$ wide, on the grid.
When a walker position is on a stripe it has a higher probability to
move 
along the stripe. The effect of such imposed anisotropy in a two-cell
system is shown in Figure \ref{2cellanisotropy}.
The anisotropy is both parallel, figure \ref{2cellanisotropy}a, and
perpendicular, figure 
\ref{2cellanisotropy}b, to the line connecting the two cells. When the
comb of stripes is 
parallel the walkers can find their path and form synaptic
connections, But, when the comb of stripes is perpendicular the walkers
cannot find their path and we can see that two out of 6 walkers lose
their way.

We constructed a computer simulation of a large network and
analyze whether the results agree with the experiments.
In figures \ref{anisotropy} we show the effect of 2-fold imposed
anisotropy  
on the growth of a large network. 
The 2-fold anisotropy has varying distances
between stripes: \ref{anisotropy}a. $5a_0$, \ref{anisotropy}b. $30a_0$,
and \ref{80normal} no stripes.  
In each figure there are 80 cells on a
$200\times200$ grid which correspond to an average
inter-cellular distance in a random cell distribution of $\sim22a_0$.
In comparison with this distance, the distance between stripes spans a
range of 
spacings from smaller to larger than the inter-cellular distance. Small 
inter-line spacing ($5a_0$) leads to strong alignment of the neurite 
with the stripes due to the high surface coverage ($33\%-20\%$) with stripes.
Large line spacing, which is beyond the average inter-cellular distance,
makes the stripes too sparse to support neurite alignment and the
produced patterns are 
similar to the one we show in figure \ref{80normal} where there are no stripes at all.
These observations are in contrast to the experimental results of Dwir
et al \cite{Dwir96}.
In their experiments small inter-line spacing reduced neurites
alignment since 
neurites can form interconnections between cell across Poly-L-Lysine.
In our model the walkers can migrate across the surface which is
uncovered by PLL with no stripes.
Therefore large line spacing reduces neurite alignment. We
expect that in an experiment in which the surface is covered with
Poly-L-Lysine and stripes are printed on it (i.e additional cover of
PLL), we will observe similar patterns to our model's simulations.

\section{The connection distance histogram}

We look for a quantitative way to compare our model results to
experiments. We suggest to use the histogram of number of connections
vs. lengths. In connections lengths we mean the straight line distance
between two connected neurons and not the length of the neurite connecting
them. We show the results in figure \ref{histogram}.

We can explain these results as follows:
In our model there are two characteristic length scales. The
inter-cellular distance $d_c$,
and the critical length for the walkers' sensitivity switch, $L_c$. 
When $L_c \sim d_c$ there is only one characteristic length scale,
hence, most of the connections' lengths are of the order of one length
and there is only one maximum in the histogram.
When $L_c > d_c$ there is another length 
scale that emerges as we show in figure \ref{histogram}b-d.
There are two maxima in the connection histogram which correspond to
the two length scales in the system.

In figure \ref{histogram}e we compare the case of chemotactic navigation
with the non-chemotactic navigation. In the first case the histogram shows
a concentrated peak at the inter-cellular distance. The histogram of the
second case is widely spread. We suggest to use these results to show the
validity of our model in experiment.

\section{Conclusions}

We have presented a navigation strategy for micro-level network
organization. This mechanism lead to the
formation of neural networks with fine structures, which can be
genetically dependent. Our results 
raise two fundamental issues: 1. One needs to develop characterization
methods (beyond number of connections per neurons) to distinguish the
various possible networks. 2. What is the relation between the network
organization and its computational properties and efficiency. 

To clarify what we mean by characterization method, we consider the
following example in which the network is mapped onto a directed graph. 
Each neuron is represented by a vertex and each
synaptic connection between neurons is represented by a directed edge
between the two 
vertices representing the neurons. We can map the graph structure into
an adjacency matrix where  
there is 1 in the i,j entry of the matrix if there is a directed edge between
the i and j vertices and 0 otherwise. Then using an algebraic
method, such as spectral theory of matrices,
we can analyze the adjacency matrix of the graph which characterizes
the network. 
In other words, just as astronomers study spectra to determine the
make-up of distance stars, we try to deduce the principle properties
and structure of the graph from its adjacency matrix spectrum
\cite{Chung97}.  

The next step of the endeavor presented here is to include the effect
of the network 
electrical activity on its self-wiring. Once some initial synaptic
connections are formed, the network begins its electrical activity. It
is natural to expect that from this point on the wiring dynamics depend
on the electrical activity. An example to such a dependence could be
that the neurons emit chemo-attractant while it bursts a train
of spikes. In addition we assume that a growth cone is more sensitive
to the chemo-attractant when its soma bursts. Using such 
navigation strategy we expect that neurons with correlated electrical
activity will have higher probability to form a synaptic connection as
we show in figure \ref{correlated}.

To conclude, we are still far from understanding the emergence of a
functioning brain from a collection of neurons. Yet we believe our
studies provide a useful first step towards this goal.

\section*{Acknowledgments}
We have benefited from fruitful discussions with E. Braun,
B. Dwir, D. Horn, D. Michaelson, I. Golding and D. Kessler. We thank
E. Braun for 
sharing with us his unpublished results. This research has been 
supported in part by a Grant from the BSF no 95-00410, a Grant from
the Israeli Science Foundation, and the Sigel Prize for Research. 

\newpage


\newpage

\section*{Figure 1}
Neuron's culture grown on a Silicon wafer after 5 days 
($500\times$ magnification, reflected light+Numarsky). 
The thin filaments are the neurites that connect the neurons.

\section*{Figure 2}  
Growth cones' strategy for path finding. a. At the beginning of the
growth the growth cone is mainly affected by the
chemo-repellent. b. After the critical length determined by the growth
cone's internal energy, the growth cone emits a quantum of triggering
agent and reverts its response to the chemotaxis.

\section*{Figure 3}
The non-uniform probability distribution of the walker.

\section*{Figure 4}
A simulation of a 50-cell system on a grid of size $100\times100$. The
small hexagons represent the neurons' soma, the filaments represent the
neurites and the dots represent the walkers.

\section*{Figure 5} 
Random walker with the non-uniform probability in figure 4. The
hexagon represents the neuron's soma and the dot represents the walker.

\section*{Figure 6}
Random walker with the ``winner take all'' rule.

\section*{Figure 7}
a. The function $G_R$ as a function of $\mbox{d}E/\mbox{d}t$.
b. The function $G_A$ as a function of $\mbox{d}E/\mbox{d}t$.

\section*{Figure 8}
Two cells system: the hexagons represent the soma, the dots represent
the walkers, and the filaments represent the neurites.

\section*{Figure 9}
The effect of chemo-attractant on the efficiency of navigation. The
cell on the right 
is a "normal" cell and the cell on the left is a ``variance'' cell that
does not emit neurites. 
Note that even a walker that first migrates away from the target cell
navigates towards this cell after it has been triggered. The contours
correspond to different concentrations of the chemo-attractant.

\section*{Figure 10}
the effect of turning off the chemotactic communication. a. When we
turn off the chemotaxis communication the walkers are unable to reach
their synaptic target. b. When we turn the communication on the walker
can migrate around the barrier.

\section*{Figure 11}
Simulations of five-cell systems. The central cell is ``normal'' and the
four cells at the corners are ``variance'' cells. The contours
represent the chemo-attractant concentration levels. a. Low rate of
``internal'' energy ``feeding'' so that \( l_c \) is much smaller than the
inter-cell distances. In this case the 
wiring is not efficient as the central cell is wired only to two of
the four neighbors. 
b. Higher rate of ``feeding'' so that \( l_c \) is approximately half
the inter-cellular 
distance. In this case the wiring is more efficient and the central
cell is wired to all its neighbor cells.

\section*{Figure 12}
Simulations of a system composed of 29 cells. Only the cell at the center
is ``normal'' and all other cells are ``variance''. The central cell has
four nearest neighbor (NN) cells and eight next nearest neighbor (NNN)
cells.  
At the beginning of the growth \( l_c \) is about half the
inter-cellular distance. Thus the central cell is wired 
only to its NN cells. After the central cell forms two connections the
``feeding'' rate doubles (doubling of 
\( l_c \) ). The new neurites navigate to the NNN cells. It demonstrates
the manner in which the cell's soma can regulate self-wiring.

\section*{Figure 13}
The effect of a saturated media with chemo-attractant on the growth process. 
The cell on the right is a ``normal'' cell and the cell on the left is a
``variance''. Since the chemo-attractant feeds the walkers,
$ \mbox{d} E/ \mbox{d} t$ remains positive during the growth, and the 
walkers do not emit the triggering agent. As a result, the walkers 
cannot find their path to the target cell.

\section*{Figure 14}
One cell and an 
artificial source of chemo-attractant: since we assume, in our model,
that at the beginning of the growth process the growth cones are
insensitive to the 
chemo-attractant, the walkers are not effected by the
chemo-attractant field.

\section*{Figure 15}
The effect of 2-fold imposed anisotropy:
we impose a comb of stripes \( a_0 \) wide and \(5a_0 \) wide between
the stripes. 
The growth cones have higher probability to move along the stripes. The
strips are parallel to the line connecting the two cells for a, and
perpendicular to that line for b. The resulting pattern agrees with
experimental observations \cite{Dwir96}.

\section*{Figure 16}
A 80-cells system on a $200\times200$ grid with a flat surface 
(no stripes).

\section*{Figure 17}
Systems of neuron growth on 2-fold anisotropy with varying distances
between stripes: a. $5a_0$, b. $30a_0$. 
In each figure there are 80 cells on
$200\times200$ grid which correspond to an average
inter-cellular distance in a random cell distribution of $\sim22a_0$.

\section*{Figure 18}
The number of connections per distance histogram. a-d The
characteristic length of the walkers' response switch varies from
$1.25a_0$ (low feeding) to $50a_0$ (high feeding). At lower feeding
rates the dominant length scale is 
is the inter-cellular distance which is $\approx20a_0$. The
connections histogram achieves its maximum at the inter-cellular
distance. At high feeding rate, there are two distinct length scales
the inter-cellular distance and characteristic length of sensitivity
switch. There are two picks at the
connection histogram which correspond to each length scale.
e. The growth cones perform random walker and establish connection
with the first soma it encounter on its way (the widely spread
histogram), the peaked histogram represent the case where $l_c=1.25a_0$.

\section*{Figure 19}
Example of the effect of the correlation in the electrical
activity. The central cell is normal, while the others are variances
which are unable to emit neurites. We choose to use variance in order
to make the connection patterns more transparent to the eye.
The central and the outer cells
(up, down, left, right) fire simultaneously while the other cells fire
at random. Indeed, as we expect the first connections are formed
between the cells that fire simultaneously.

\newpage

\end{document}